# Single-Photon-Assisted Two-Photon Polymerization

Buse Unlu[1,*], Maria Isabel Álvarez-Castaño[1], Antoine Boniface[1,2], Ye Pu[1], Christophe Moser[1]

[1]Laboratory of Applied Photonics Devices, School of Engineering, Ecole Polytechnique Fédérale de Lausanne, CH-1015, Lausanne, Switzerland
[2]AMS Osram, Martigny, Switzerland

Corresponding author: buse.unlu@epfl.ch

**Abstract**

Light-based additive manufacturing (AM) has revolutionized the fabrication of complex three-dimensional (3D) objects offering a cost-effective and high-speed alternative to traditional machining. One-photon polymerization is a key process in this advancement, standing out for rapid printing time, albeit with limited resolution. Two-photon polymerization (2PP) empowers AM with unprecedented resolution but is accompanied by a tradeoff of prolonged printing times.
We propose combining the single-photon absorption (1PA) and 2PP to benefit from the dual capabilities, allowing for faster printing while maintaining high resolution and improved depth sectioning, respectively. In this study, we employ a blue light source to pre-excite a photocurable resin by 1PA followed by a precisely focused femtosecond (fs) beam to provide the missing energy necessary to reach the polymerization threshold to solidify the resin through two-photon absorption. First, we investigate the impact of pre-sensitization by blue light illumination on 2PP and demonstrate one order of magnitude faster printing time for a voxel size of 150 nm as compared to the same voxel size printed by 2PP only. Then, we build a custom 2PP printer utilizing blue light sensitization in a light-sheet mode and demonstrate successful 3D prints.

**Introduction**

Additive Manufacturing (AM), enables fabricating complex objects which are challenging to produce using conventional machining techniques, at a low cost and high-speed [1,2]. It significantly reduces or eliminates assembly time and supply chain delays, allowing for on-site production. Photopolymerization is one of the most popular techniques used in light-based AM, and is also the first 3D printing method (stereolithography, invented by Hideo Kodama in the early 1980s [3]). In this process, a photocurable resin comprising monomers (and/or oligomers), inhibitors and a photoinitiator (PI) undergoes a chemical reaction upon irradiation by visible or ultraviolet (UV) light [1,4] which solidifies the liquid resin to a solid upon phase change. Hardening of the material is only induced when a sufficient portion of cross-linked bonds is formed. This occurs when the light dose corresponding to the light energy deposited over a given exposure time, crosses the polymerization threshold. The activation of the photoinitiator that creates a radical at the origin of the polymerization process can be done either via a single- or two-photon absorption (2PA) [5].

Single-photon absorption (1PA) is linearly proportional to the light intensity at which a photoinitiator molecule absorbs one-photon. A typical wavelength for photoinitiator absorption is in the ultraviolet-violet range (300-400 nm) [6]. Due to the high quantum efficiency of PIs, the polymerization and thus the printing speed can reach the order of $10^6$ voxels/s voxel printing rate [7]. However, as the absorption and thus polymerization is proportional to intensity, the resin polymerization occurs along the entire axial extent of the focused laser beam which prevents obtaining depth sectioning, leading to poor axial resolution and limiting the printing of complex 3D parts with small features in all dimensions. Stereolithography (SLA) [8] uses a 1PA process and goes around this depth sectioning problem by loading the resin with absorbers to restrict light penetration deep inside the resin [9]. In this case, the axial resolution is determined by the decay of light due to absorption. However, it is not energy efficient as this light is lost as heat in the material. Instead of the layer-by-layer solidification as in SLA, Tomographic volumetric additive manufacturing (TVAM) offers to print the entire volume at once, by projecting 2D images (computed from the object 3D geometry) with a digital light projector from multiple angles by rotating the resin cuvette. The cumulative dose produced by all projections at different angles results in all targeted voxels surpassing the polymerization threshold limit and solidifying the liquid resin in the desired geometry [10,11]. In TVAM, the light absorption by resin needs to be low to ensure that light crosses the entire volume. Thus, this technique provides a rapid printing time of the order of a few tens of seconds for centimeter-scale objects [12]. The resolution is limited so far to 50-80 µm for cm scale objects and down to 20-30 µm for mm scale objects [13].

In 2PA, the molecule absorbs two photons of half-energy simultaneously where the sum of the energy is equal to the total transition energy [14]. The feature distinguishing 2PA from 1PA is the quadratic dependence of absorption with the incident light intensity. This offers axial sectioning at the focal plane and thus enables the printing of three-dimensional (3D) structures by scanning the focal spot in 3D. Feature size down to around 100 nm can be achieved [15–20]. A high light intensity is essential in 2PA to obtain appreciable absorption due to the low 2PA cross-section [14,21]. Thus, femtosecond (fs) pulsed lasers are used in practice [22]. Several techniques have been studied to accelerate the printing process, namely two-photon grayscale lithography (2GL) [23], spatiotemporal focusing [24] and multi-focus printing [7,25–27]. For instance, work from the Nieder group has sped up the fabrication by 9 times using a diffraction optical element (DOE) operated to split the single pulses IR laser beam into several beams [25]. A printing speed of $10^7$ voxels/s using this approach has been reported [7]. Nevertheless, the multi-focus method has limitations due to the restricted field-of-view imposed by the high numerical aperture (NA) microscope objective, possible interference and distortion that may result from the short distance between foci, and mainly the need to increase the laser power proportionally with the number of laser foci. There are new reports of an effective two-photon effect by a so-called 2 step absorption that uses continuous wave low-power (around 100 µW) sources [28].

Recent work by the Brunner group has employed both single-photon polymerization and 2PP sequentially during the fabrication of waveguides to reduce the 2PP printing time [29]. Essentially, 2PP is used where high resolution is needed and single-photon flood illumination where low resolution is sufficient. They report a printing time reduction of approximately 90% compared to printing with only 2PP, as the large volume around the small waveguides printed using a fs laser 2PP is cured by UV illumination.

Inspired by the dual use of 1 PA and 2PA, here, we propose a printing method that takes advantage of the dual use of single-photon polymerization and two-photon polymerization to speed up the printing process while maintaining the high resolution and better depth sectioning that 2PP offers through nonlinearity and strongly localized volume.

We address the impact of pre-sensitization of a photocurable resin by single-photon absorption on the speed and resolution of 2PP. Because the 1PA process is very efficient, the onset of polymerization could be shortened as compared to using only 2PA. In what follows, we report on a considerable acceleration in printing time by first sensitizing a photosensitive resin with a blue light continuous beam prior to focusing a femtosecond beam, achieving a minimum lateral resolution of 150 nm 10 times faster than when only using a femtosecond beam. Then, we use this result to build a 2-photon printer whereby blue light sensitization is used in a light-sheet mode and a femtosecond laser is focused and scanned to build a 3D object. We report that the blue light-sheet-assisted two-photon polymerization speeds up the initiation of polymerization, and enhances the quality of the printed surface.

**Results**

*1. Blue Light Illumination Assisted Two-Photon Polymerization*

We propose to investigate the effect of single-photon sensitization followed by two-photon excitation. In particular, we characterize two relevant resin properties: the threshold dose required to polymerize the photocurable resin and the resulting printing resolution. For this purpose, blue light at 405 nm first irradiates the resin formulation and then, a fs laser beam at 780 nm is focused on the resin to form a polymerized voxel.

The set-up is illustrated in Figure 1.a (a comprehensive sketch of the experimental setup is depicted in the Supplementary material Fig. S.1). We use a Mode-locked femtosecond (fs) pulsed Ti:Sapphire laser operating at a center wavelength of 780 nm with a repetition rate of 80 MHz and 70 fs pulse duration to induce photopolymerization through 2PA. A phase-only Spatial Light Modulator (SLM, PLUTO-2.1 by Holoeye) is used to laterally scan the fs laser beam over the printing area by varying the period of a phase grating displayed on the SLM. A microscope objective (100x/1.25 OIL) focuses the fs laser beam. The specular reflection of the focused fs beam, from the cover glass substrate (thickness of 0.17 mm) containing the resin, is directed by a dichroic mirror and imaged to a camera. This imaging system is used to calibrate the position of the focused beam with respect to the printing surface as well as the shape of the focused beam. The glass substrate is coated with 3-(Trimethoxysilyl)propyl methacrylate to enhance the adhesion of the printed structures [30,31]. A blue light illumination system consisting of a collimated fiber coupled single-mode 405 nm laser diode is integrated into this 2PP setup. An optical beam shutter is mounted on the beam path to control the exposure time of the blue light illumination.

Numerous commercially available photoinitiators (PIs) are utilized in literature for stimulating polymer chain formation through either single-photon or multiphoton absorption. Lucirin TPO-L (ethyl- 2,4,6trimethylbenzoylphenylphosphinate) is one of the PIs existing in liquid form which has high solubility and compatibility with several resin mixtures. Despite TPO-L being designated as a single-photon absorption PI, it has also been utilized in 2PP [30,32–34].

Contrary to other PIs, TPO-L has a high radical quantum yield of 0.99 for 1PA in the range of 350 nm to 425 nm [30,34] and also has 2PA [30,32]. In line with our objective, TPO-L fulfils the requirement for a material having both single- and two-photon absorption. In this study, a photocurable organic acrylate-based resin (Sartomer, PRO21905) has been formulated for printing, with TPO-L as the photoinitiator (concentration 2.75 wt%).

The photocurable resin is sandwiched between two cover glasses (thickness of 0.17 mm). The response of the resin to the fs illumination is first characterized. An array of focused spots is printed by varying the average fs light power: 10,11, and 15 mW and the exposure time of the beam spots is varied (from 50 ms to 3 s) along each row of the array. The experiment is repeated 3 times. The minimum exposure time (50 ms) is limited by the response time of the SLM (see SLM characterization in the Supplementary material Fig. S.2). After the printing process, the unpolymerized resin surrounding the printed voxels is developed (rinsing) in propylene-glycol-monomethyl-ether-acetate (PGMEA) followed by isopropanol (IPA). Then, we investigate the size of the polymerized voxels by imaging the polymerized voxel with a differential interference contrast (DIC) microscope. To examine voxel sizes below 300 nm, a scanning electron microscope (SEM) is utilized. An example of a printed array image taken with DIC is shown in the inset of Figure 1.a. Average diameters of the printed voxels are measured and plotted as a function of the applied light dose as shown in Figure 1.b-c, which are fitted with a logarithm function. Printing with average fs light power of 20,23,25, and 30 mW is also characterized in the Supplementary material Fig. S.4. As expected, the printed voxel size decreases when the average power and the exposure time are reduced. A lateral resolution of 150 nm is achieved with a dose of $1.84 \times 10^8$ mJ/cm$^2$ (corresponding to 10 mW fs laser average power and exposure time of 500 ms) in Figure 1.b.

We then proceed with the characterization of the resin response to both blue light illumination and fs illumination. The blue light exposure dose is calibrated so as to be below the polymerization threshold (see Supplementary material Fig. S.3). The dual illumination for 1PA and 2PA is then applied: the resin is first illuminated by a blue light dose at 3.1 mJ/cm$^2$ (corresponding to an illumination power of 12 mW and 50 ms exposure time) over an area of 0.193 cm$^2$, and then irradiated with 10 mW fs laser average power. As shown in Figure 1.b, the voxel size is considerably larger compared to the case when only the fs beam is employed. This indicates that blue light sensitization influences the dynamics of polymerization when it is followed by fs irradiation. Quantitatively, at the exposure dose of $1.84 \times 10^8$ mJ/cm$^2$, the printed voxel size is 7 times larger than when using only fs irradiation.

It is interesting to note that the voxel growth using both 1PA followed by 2PA is fitted using a sum of two logarithms whereas a single logarithm is enough when only 2PA is used. This indicates that there are two times scales when there is both 1PA and 2PA.

Conversely, one needs less average fs power to obtain a given voxel size when the resin is pre-irradiated by blue light. This is illustrated in Figure 1.c: a voxel size of 1.5 $\mu$m is obtained using 7 mW of fs power (with blue light dose at 3.7 mJ/cm$^2$) compared to the same voxel size achieved with more than twice (15 mW) fs power (without blue light). A voxel size of 150 nm resolution is also achieved when the polymerization is assisted by blue light illumination. Here, the fs power is reduced from 10 mW to 7 mW and the exposure time is decreased by 10 times (from 500 ms to 50 ms) compared to the case where polymerization is only obtained via 2PA.

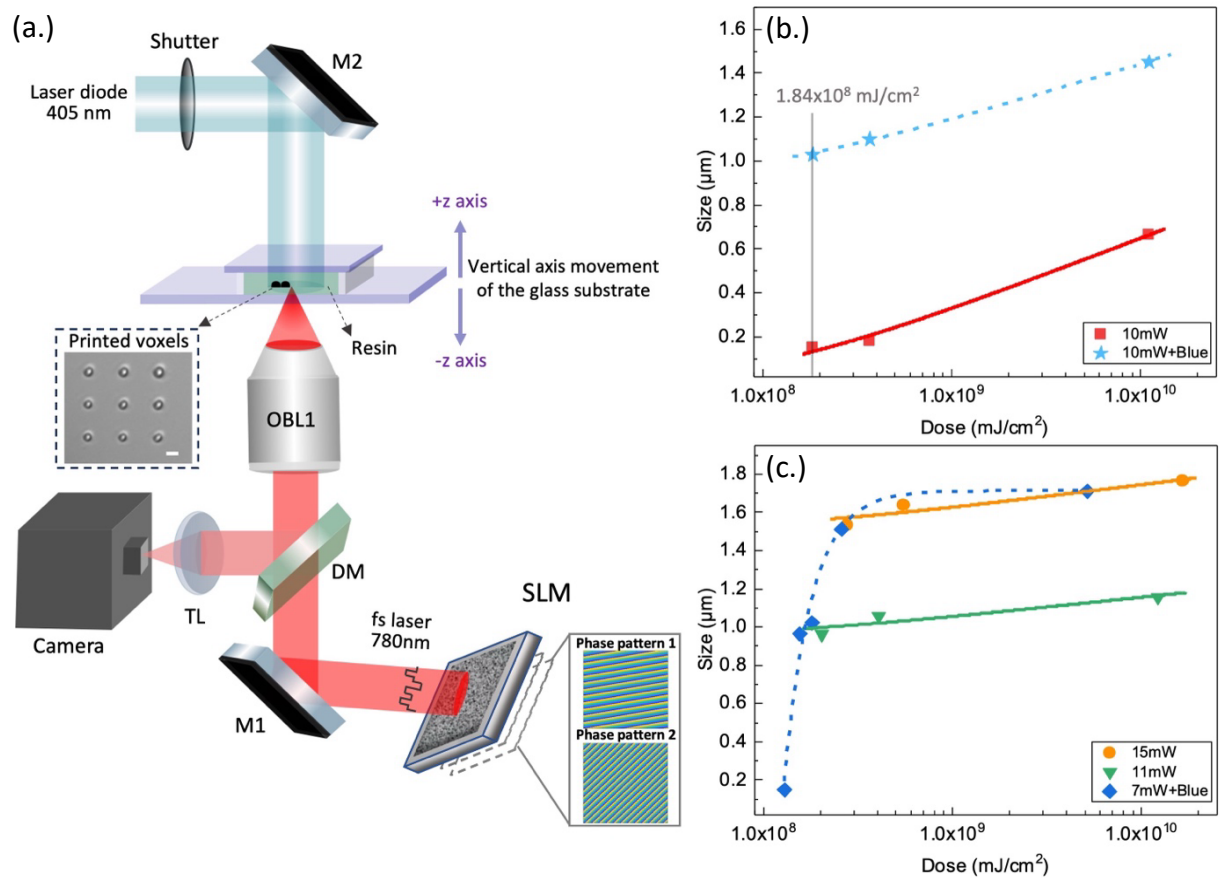

**Figure 1.** Blue light illumination assisted two-photon polymerization. (a.) Schematic illustration of the experimental setup. M: mirror, OBL1: objective lens 100x/1.25, TL: tube lens, DM: dichroic mirror. Inset: DIC image of printed voxels by 15 mW fs laser average power. The scale bar is 2 µm. (b.) Dose test with fs laser power of 10 mW with and without blue light irradiation of 3.1 mJ/cm$^2$. (c.) Dose test with fs laser powers of 15 mW, 11 mW and 7 mW assisted by blue light illumination of 3.7 mJ/cm$^2$. A logarithmic function is used to fit data points.

*2. Blue Light-Sheet Assisted Two-Photon Polymerization for Printing 3D Structures*

For 3D object printing, we implement a blue light-sheet resulting in a single-photon absorption sheet to assist 2PP and thus providing a pre-sensitizing effect for an earlier onset of polymerization by 2PA. Here, the fs laser beam is focused within the light-sheet thickness. The printing parameters are optimized such that the resin only polymerizes at the intersection volume of the two light sources.

Figure 2.a illustrates the blue light-sheet assisted 2PP setup, which is detailed in Supplementary material Fig. S.5. A cylindrical lens is placed on the light path of the 405 nm laser diode to form a light-sheet for 1PA. The light is directed through a low NA objective lens (4x/0.10) that is mounted on a translation stage, which allows for the precise adjustment of the light-sheet waist position. First, the blue light-sheet is characterized as demonstrated in Supplementary Fig. S.6. It has a beam waist of 3.4 µm (5.02 µm) at the focal region and a Rayleigh range of 89.7 µm (289 µm) in air (respectively, inside the resin, refractive index =

1.48). The light-sheet is focused inside a cuvette containing the photocurable resin. Orthogonal to it, the focus spot of the fs laser at 780 nm is formed by a low NA objective lens (20x/0.45) and scanned by an SLM within the FOV of the objective lens. A glass rod with a 700 μm diameter is employed as a printing substrate, on which a coated coverslip (700x700 μm$^2$) is glued to ensure adherence of the printed parts to the substrate. The focal point of the fs laser is aligned to be within the waist of the blue light-sheet which lies at the surface of the rod. The rod is dipped into the resin container as shown in Figure 2.b.

A thickness of the resin equal to the light sheet thickness undergoes pre-sensitization via 1PA by the blue light-sheet which delivers a dose below the polymerization threshold. Next, the focused fs laser spot exposes the resin within the light-sheet region. Polymerization is initiated on the surface of the glass rod, and a 3D structure can be formed point-by-point by digitally scanning the fs-focused spot on a xy plane and layer-by-layer by moving the rod along the z-axis. The printing process is monitored by the camera depicted in Figure 2.a. A red LED provides the illumination through the glass rod.

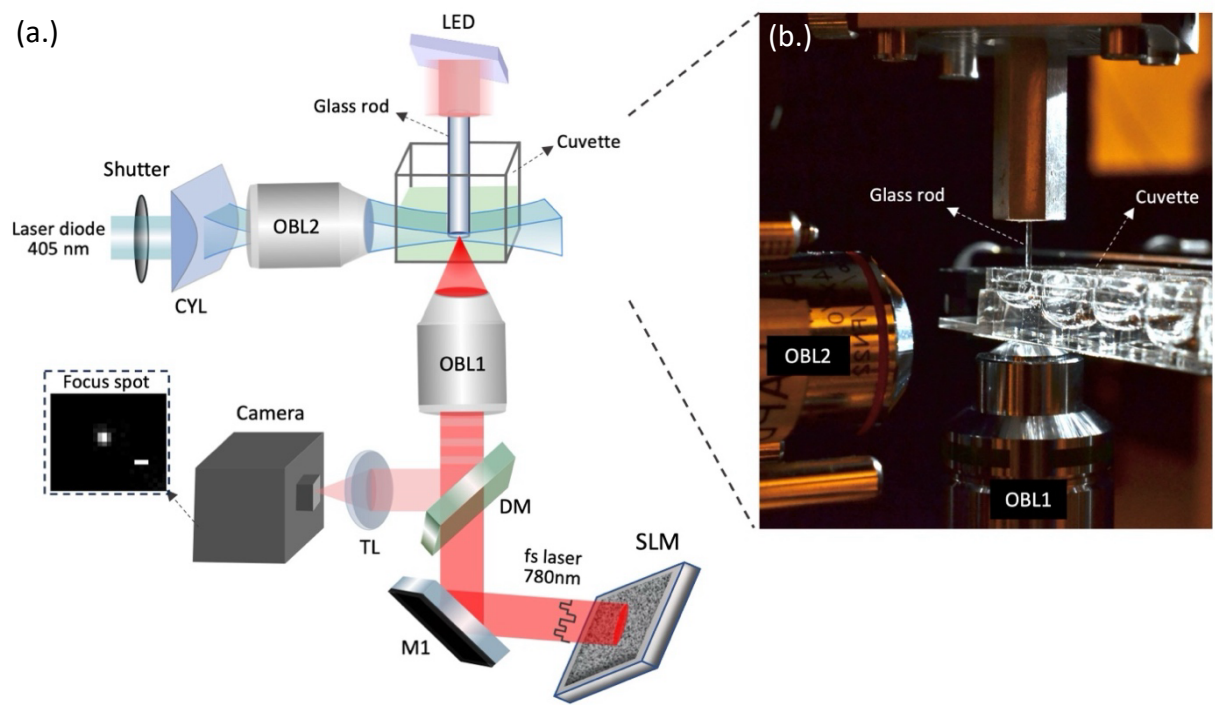

**Figure 2.** Blue light-sheet assisted two-photon polymerization for printing 3D structures. (a.) Schematic illustration of the experimental setup. M: mirror, DM: dichroic mirror, TL: tube lens, CYL: cylindrical lens, OBL1: objective lens 20x/0.45, OBL2: objective lens 4x/0.10. The inset shows the fs laser focus spot reflected from the glass rod dipped in the resin. The scale bar is 2 μm. (b.) A real photograph of the printing region in the setup where the focus of the fs laser (OBL1) coincides with the blue light-sheet (OBL2) on the tip of the glass rod dipped into the resin container.

In the experiment, first, a building-like structure (tall rectangular prism) is used as the model to investigate the effect of blue light sensitization on voxel size and thus printing time. For all printing trials of the rectangular prism, the hatching distance (distance between the voxels, lateral overlap) is chosen to be 760 nm along the x-axis and 430 nm along the y-axis. The slicing distance (distance between two consecutive layers) is first set at 3 μm nearly equal to the light sheet thickness. Before the blue light-sheet exposure, the printing parameters for 2PP are characterized: a focused fs laser spot of 1.4 μm full-width half maximum (FWHM) diameter (in the inset of Figure 2.a) with average power varying from 14 mW to 20 mW in 2 mW increments with a 50 ms exposure time is applied to irradiate the resin. None of the power levels, except 20 mW, yielded a printed structure. The 3D structure (30 layers) printed with 20 mW average fs laser power and 50 ms exposure time is depicted in Figure 3.a. As it can be observed in the zoomed region (inset of Figure 3.a), the hatching distance in the x direction and slicing distance prevents the printed voxels from merging well with each other. To investigate the effect of the resin's pre-sensitization on the printed structure, the experiment is repeated under the blue light-sheet illumination with an average power of 0.3 mW and an exposure time of 50 ms (given in Supplementary material Fig. S.3), before the fs laser illumination. We observe a larger voxel size that fills well the gap between the printed voxels resulting in a better visual surface quality as shown in Figure 3.b. To achieve the 3D structure with the same surface quality without blue light-sheet sensitization required 28 mW of fs laser power with an exposure time of 100 ms as shown in Figure 3.c. Due to the sensitization effect of the blue light-sheet, the fs laser power was reduced by 1.4 times (from 28 mW to 20 mW) and the exposure time by 2 times (from 100 ms to 50 ms) compared to polymerization performed only by 2PA.

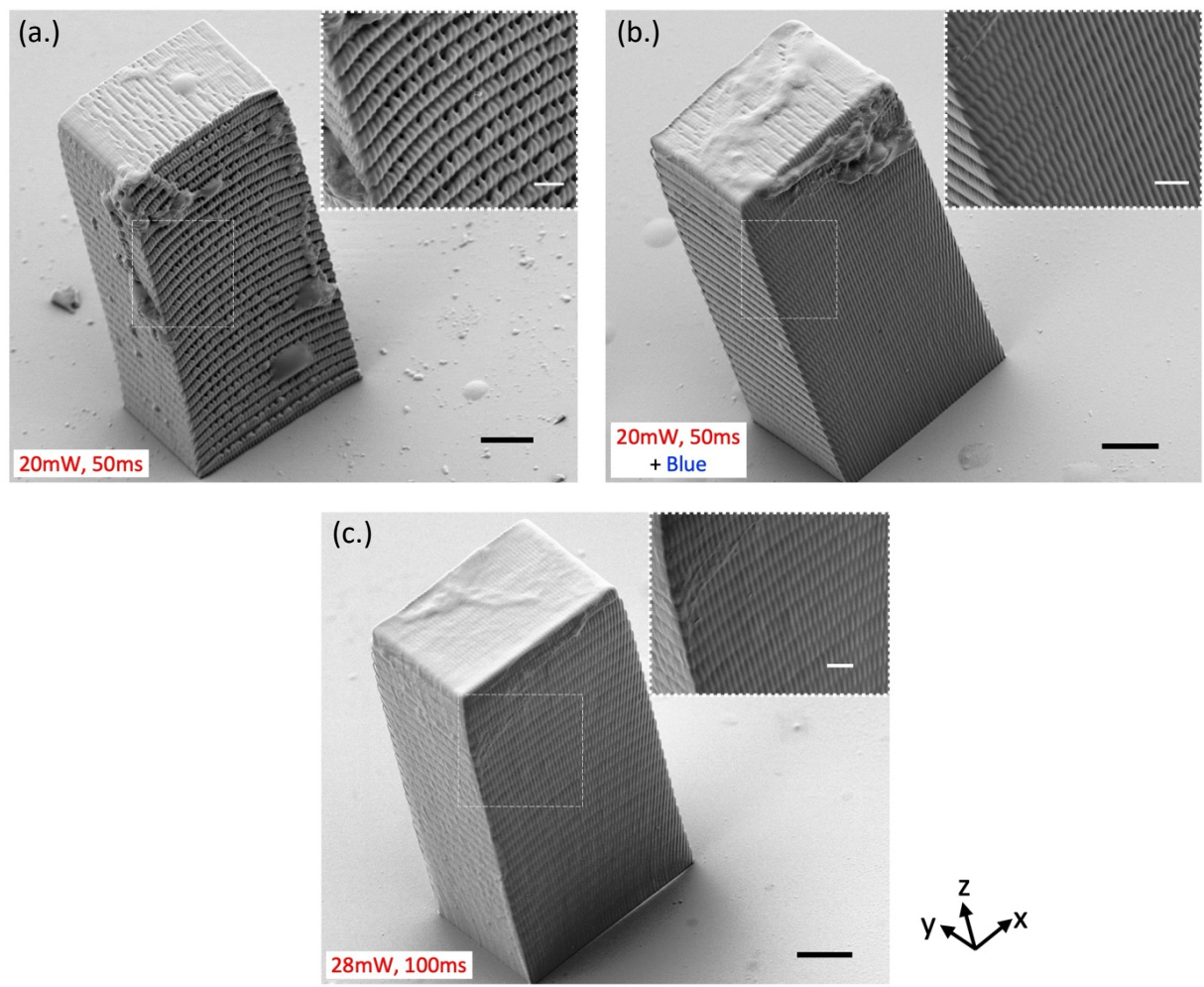

**Figure 3.** SEM images of the printed 3D building-like structure (a.) only by a fs laser power of 20 mW and exposure time of 50 ms (b.) with the blue light-sheet sensitization, and (c.) only by a fs laser power of 28 mW and exposure time of 100 ms without the blue light-sheet sensitization. Insets show a zoom-in to the areas indicated by the dashed rectangles. The scale bars are 10 μm and 3 μm for the insets.

Since power levels lower than 20 mW result in shorter voxel heights, we reduced the slicing distance from 3 µm to 1.5 µm, facilitating the attachment of layers to each other at the lower power levels. Figure 4.a shows the computer-aided designed (CAD) model of the structure. Hatching distance is set to 760 nm along x- and y-axes. We used fs laser average power of 17 mW with an exposure time of 50 ms. Printing was achieved only when the resin was sensitized by the blue light-sheet (average power of 0.3 mW and exposure time of 50 ms) before the fs laser irradiation. Otherwise, the fs laser dose was insufficient to print. The SEM image of the printed structure is depicted in Figure 4.b.

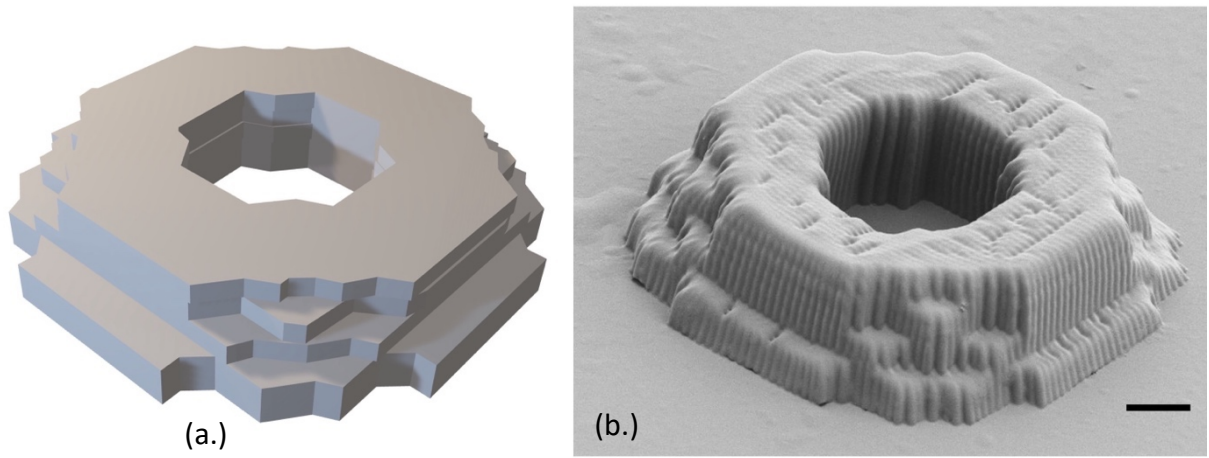

(a.)         (b.)

**Figure 4.** (a.) 3D CAD model of a structure. (b.) SEM image of the printed structure with a scale bar of 5 µm.

*3. Depth resolution*

In the previous experiments with the assistance of the blue light-sheet, the depth resolution is given by the light sheet thickness. Here, we show that the depth resolution can be shorter (at least by a factor 2) than the light sheet thickness.

We use the SLM to scan the beam with the light sheet thickness rather than moving the rod to avoid the motion of the resin. This is implemented by adding a Fresnel mask on the SLM pattern. By adjusting the quadratic phase on the SLM (detailed in Supplementary material 7) the focus spot is moved along the axial direction.
We tested printing two layers within the single light-sheet thickness using an average power of 17 mW for the fs laser and 0.3 mW for the blue light-sheet sensitization with an exposure time of 50 ms for both illuminations. As illustrated in the printing flow chart in Figure 5.a, the resin is first exposed to the blue light-sheet and then the fs laser is focused on the first half of the 3 µm light sheet thickness to print the first layer. Subsequently, the focus of the fs laser beam is digitally axially moved by 1.5 µm within the already sensitized region using the Fresnel lens on the SLM. To print the third layer, the rod is moved up in z-direction, and the same procedure is repeated 5 times. The 3D structure (10 layers) obtained is depicted in Figure 5.b. This demonstrates that layers of voxels can be printed within a given light-sheet.

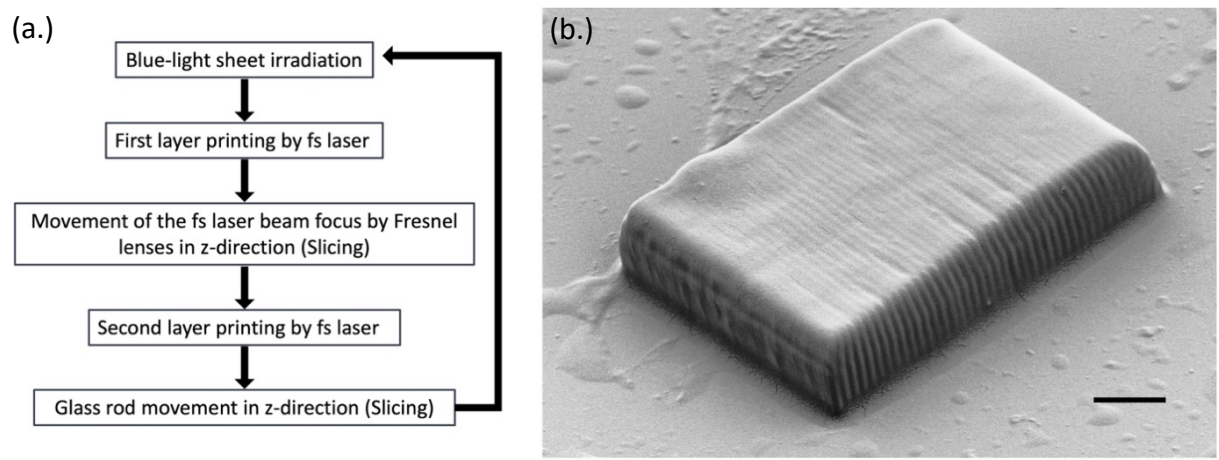

**Figure 5.** Digital three-dimensional control in printing by Fresnel lenses. (a.) Printing flow chart. (b.) SEM image of the printed 3D structure using Fresnel lenses. The scale bar is 5 μm.

**Conclusion**

In summary, our investigation of the combined use of single-photon sensitization followed by 2PP demonstrates a remarkable reduction in fabrication times while maintaining precision compared to only the 2PP process.

First, the experiments with blue light illumination followed by fs laser beam scanning confirm that small voxel size can be obtained with reduced fs power and exposure time, making the process several times faster than traditional only 2PP. We achieved a minimum lateral resolution of 150 nm 10 times faster with blue light sensitization.

A custom 2-photon printer with a blue light-sheet was built. Due to the sensitization effect of the blue light-sheet, we demonstrated a factor 2.8 reduction in light dose to print a comparable cuboid by 2PA. We also showed that the depth selectivity can be shorter than the light sheet thickness by at least a factor 2.

We envision that the combination is 1 PA and 2PA is well suited for TVAM, whereby all voxels of a latent object are below the hardening threshold and thus with the addition of a small dose delivered via 2PA a higher resolution object could be formed at a much higher speed.

**Disclosures**. The authors declare no conflicts of interest.

**References**

[1] Gibson I, Rosen D, Stucker B, Khorasani M. Additive Manufacturing Technologies. Third edition. Cham: Springer International Publishing; 2021. https://doi.org/10.1007/978-3-030-56127-7.
[2] Wong KV, Hernandez A. A Review of Additive Manufacturing. ISRN Mech Eng 2012;2012:1–10. https://doi.org/10.5402/2012/208760.
[3] Kodama H. Automatic method for fabricating a three-dimensional plastic model with photo-hardening polymer. Rev Sci Instrum 1981;52:1770–3. https://doi.org/10.1063/1.1136492.
[4] Saleem Hashmi, Gilmar Ferreira Batalha, Chester J. Van Tyne, Bekir Yilbas. Comprehensive Materials Processing. vol. 1. Elsevier; 2014.
[5] Paul Francis Jacobs. Rapid Prototyping & Manufacturing: Fundamentals of Stereolithography. Society of Manufacturing Engineers; 1992.

[6] Bagheri A, Jin J. Photopolymerization in 3D Printing. ACS Appl Polym Mater 2019;1:593–611. https://doi.org/10.1021/acsapm.8b00165.

[7] Hahn V, Kiefer P, Frenzel T, Qu J, Blasco E, Barner-Kowollik C, et al. Rapid Assembly of Small Materials Building Blocks (Voxels) into Large Functional 3D Metamaterials. Adv Funct Mater 2020;30:1907795. https://doi.org/10.1002/adfm.201907795.

[8] Hull CW. Apparatus for production of three-dimensional objects by stereolithography. 4,575,330, 1986.

[9] Gao Y, Xu L, Zhao Y, You Z, Guan Q. 3D printing preview for stereo-lithography based on photopolymerization kinetic models. Bioact Mater 2020;5:798–807. https://doi.org/10.1016/j.bioactmat.2020.05.006.

[10] Kelly BE, Bhattacharya I, Heidari H, Shusteff M, Spadaccini CM, Taylor HK. Volumetric additive manufacturing via tomographic reconstruction. Science 2019;363:1075–9. https://doi.org/10.1126/science.aau7114.

[11] Ligon SC, Husár B, Wutzel H, Holman R, Liska R. Strategies to Reduce Oxygen Inhibition in Photoinduced Polymerization. Chem Rev 2014;114:557–89. https://doi.org/10.1021/cr3005197.

[12] Bernal PN, Delrot P, Loterie D, Li Y, Malda J, Moser C, et al. Volumetric Bioprinting of Complex Living-Tissue Constructs within Seconds. Adv Mater 2019;31:1904209. https://doi.org/10.1002/adma.201904209.

[13] Toombs JT, Luitz M, Cook CC, Jenne S, Li CC, Rapp BE, et al. Volumetric additive manufacturing of silica glass with microscale computed axial lithography. Science 2022;376:308–12. https://doi.org/10.1126/science.abm6459.

[14] Papagiakoumou E, Ronzitti E, Emiliani V. Scanless two-photon excitation with temporal focusing. Nat Methods 2020;17:571–81. https://doi.org/10.1038/s41592-020-0795-y.

[15] Robert KA, Thompson MB. Finer features for functional microdevices. Nature 2001;412:698–9. https://doi.org/10.1038/35089135.

[16] Zheng L, Kurselis K, El-Tamer A, Hinze U, Reinhardt C, Overmeyer L, et al. Nanofabrication of High-Resolution Periodic Structures with a Gap Size Below 100 nm by Two-Photon Polymerization. Nanoscale Res Lett 2019;14:134. https://doi.org/10.1186/s11671-019-2955-5.

[17] Faraji Rad Z, Prewett PD, Davies GJ. High-resolution two-photon polymerization: the most versatile technique for the fabrication of microneedle arrays. Microsyst Nanoeng 2021;7:71. https://doi.org/10.1038/s41378-021-00298-3.

[18] Paz VF, Emons M, Obata K, Ovsianikov A, Peterhänsel S, Frenner K, et al. Development of functional sub-100 nm structures with 3D two-photon polymerization technique and optical methods for characterization. J Laser Appl 2012;24:042004. https://doi.org/10.2351/1.4712151.

[19] Gonzalez-Hernandez D, Varapnickas S, Bertoncini A, Liberale C, Malinauskas M. Micro-Optics 3D Printed via Multi-Photon Laser Lithography. Adv Opt Mater 2023;11:2201701. https://doi.org/10.1002/adom.202201701.

[20] Cao W, Yu W, Xiao Z, Qi D, Wang Z, Xin W, et al. Water repellence of biomimetic structures fabricated via femtosecond laser direct writing. J Manuf Process 2023;102:644–53. https://doi.org/10.1016/j.jmapro.2023.07.076.

[21] Göppert-Mayer M. Elementary processes with two quantum transitions. Ann Phys 2009;521:466–79. https://doi.org/10.1002/andp.200952107-804.

[22] Harinarayana V, Shin YC. Two-photon lithography for three-dimensional fabrication in micro/nanoscale regime: A comprehensive review. Opt Laser Technol 2021;142:107180. https://doi.org/10.1016/j.optlastec.2021.107180.

[23] Aderneuer T, Fernández O, Ferrini R. Two-photon grayscale lithography for free-form micro-optical arrays. Opt Express 2021;29:39511. https://doi.org/10.1364/OE.440251.

[24] Somers P, Liang Z, Johnson JE, Boudouris BW, Pan L, Xu X. Rapid, continuous projection multi-photon 3D printing enabled by spatiotemporal focusing of femtosecond pulses. Light Sci Appl 2021;10:199. https://doi.org/10.1038/s41377-021-00645-z.

[25] Maibohm C, Silvestre OF, Borme J, Sinou M, Heggarty K, Nieder JB. Multi-beam two-photon polymerization for fast large area 3D periodic structure fabrication for bioapplications. Sci Rep 2020;10:8740. https://doi.org/10.1038/s41598-020-64955-9.

[26] Geng Q, Wang D, Chen P, Chen S-C. Ultrafast multi-focus 3-D nano-fabrication based on two-photon polymerization. Nat Commun 2019;10:2179. https://doi.org/10.1038/s41467-019-10249-2.

[27] Kato J, Takeyasu N, Adachi Y, Sun H-B, Kawata S. Multiple-spot parallel processing for laser micronanofabrication. Appl Phys Lett 2005;86:044102. https://doi.org/10.1063/1.1855404.

[28] Hahn V, Messer T, Bojanowski NM, Curticean ER, Wacker I, Schröder RR, et al. Two-step absorption instead of two-photon absorption in 3D nanoprinting. Nat Photonics 2021;15:932–8. https://doi.org/10.1038/s41566-021-00906-8.

[29] Grabulosa A, Moughames J, Porte X, Brunner D. Combining one and two photon polymerization for accelerated high performance (3 + 1)D photonic integration. Nanophotonics 2022;11:1591–601. https://doi.org/10.1515/nanoph-2021-0733.

[30] Baldacchini T, LaFratta CN, Farrer RA, Teich MC, Saleh BEA, Naughton MJ, et al. Acrylic-based resin with favorable properties for three-dimensional two-photon polymerization. J Appl Phys 2004;95:6072–6. https://doi.org/10.1063/1.1728296.

[31] Maruyama T, Hirata H, Furukawa T, Maruo S. Multi-material microstereolithography using a palette with multicolor photocurable resins. Opt Mater Express 2020;10:2522. https://doi.org/10.1364/OME.401810.

[32] Mendonca CR, Correa DS, Baldacchini T, Tayalia P, Mazur E. Two-photon absorption spectrum of the photoinitiator Lucirin TPO-L. Appl Phys A 2008;90:633–6. https://doi.org/10.1007/s00339-007-4367-0.

[33] Ummethala G, Jaiswal A, Chaudhary RP, Hawal S, Saxena S, Shukla S. Localized polymerization using single photon photoinitiators in two-photon process for fabricating subwavelength structures. Polymer 2017;117:364–9. https://doi.org/10.1016/j.polymer.2017.04.039.

[34] Uppal N, Shiakolas PS. Process Sensitivity Analysis and Resolution Prediction for the Two Photon Polymerization of Micro/Nano Structures. J Manuf Sci Eng 2009;131:051018. https://doi.org/10.1115/1.4000097.

# Single-Photon-Assisted Two-Photon Polymerization:

# Supplementary Materials

Buse Unlu[1,*], Maria Isabel Álvarez-Castaño[1], Antoine Boniface[1,2], Ye Pu[1], Christophe Moser[1]

[1]Laboratory of Applied Photonics Devices, School of Engineering, Ecole Polytechnique Fédérale de Lausanne, CH-1015, Lausanne, Switzerland

[2]AMS Osram, Martigny, Switzerland

### 1. Experimental setup for blue light illumination-assisted two-photon polymerization

The light illumination-assisted two-photon polymerization setup is illustrated in Fig. S.1. The light (represented by red color for 2PP) from an ultrafast Ti:sapphire laser (Mai Tai eHP DS) operating at 780 nm wavelength with 80 MHz repetition rate and 70 fs pulse width is directed, using two mirrors (M1 and M2, Thorlabs BB1-E03) and an expansion 4f optical system (L1: Thorlabs AC254-035-B-ML f = 35mm and L2: Thorlabs AC254-150-B-ML f = 150 mm), to fill the aperture of a phase-only spatial light modulator (SLM, PLUTO-2.1 by Holoeye). The laser power is adjusted by a combination of a half-waveplate (HWP, Thorlabs WPH10M-808) which can be rotated by a computer-controlled motor (shifting the polarization of the light beam causing phase delay, i.e., retardation, in a range of $\pi$) and a polarizing beam splitter (PBS, CM1-PBS252). The horizontal light polarization after the PBS is oriented to match the polarization of the SLM (PLUTO Holoeye). The Liquid Crystal-on-Silicon (LCoS) reflective mode SLM is a phase-only modulator (1920 pixels x 1080 pixels resolution, 8bit grey levels and 60 Hz input frame rate). When a phase grating is loaded onto the SLM, the incident input light is diffracted into many orders. The first-order is allowed to pass through an iris located in the Fourier plane of the first lens (L3, Thorlabs LA1229-B-N-BK7 f = 175 mm) placed after the SLM and the unmodulated zero-order is blocked by it. The angle of the laser beam is given by the first diffraction order, which is controlled by the phase grating period loaded onto the SLM. As depicted in Fig. 1.a, by varying the period of the phase patterns applied to the SLM, it controls the plane wave angle. A 4F system by L3 and L4 (Thorlabs LA1229-B-N-BK7 f = 175 mm) images the phase pattern of the SLM on the back focal plane of the focusing high numerical aperture (NA) objective lens (OBJ1, Leica C-Plan 100x OIL NA = 1.25). This creates a focal spot, after the objective, that can be translated within the field of view (FOV). The combination of a dichroic mirror (DM, Semrock FF699-FDi01-t1-25x36), tube lens (TL, Thorlabs TRH254-040-A-ML f = 40 mm) and camera (Basler acA2040-55um) are mounted to image the specular reflection of the focused fs beam, from the cover glass substrate (EMS #1 1/2) containing the resin. A neutral-density filter is also used in front of the camera to attenuate the power of the beam to prevent camera saturation.

A blue light illumination system is integrated into the 2PP setup. A laser diode working at 405 nm wavelength (Thorlabs, L405P150) is coupled into single-mode fiber (SMF, Thorlabs P3-405B-APC) using an aspheric lens (L5, Thorlabs 354330-A f = 3.1 mm) and a fiber collimator (FC, Thorlabs F671APC-405 f = 4.02 mm), and its divergent output is converted into a parallel beam of light by an aspheric lens (L6, Thorlabs C230TMD-A f = 4.51 mm). A beam splitter (BS, Thorlabs BS031 50:50) is used on the light path to measure the incidence power. An optical beam shutter is mounted on the light path to control the exposure time of blue light illumination that is directed to the resin by a mirror (M7, Thorlabs BB1-E01).

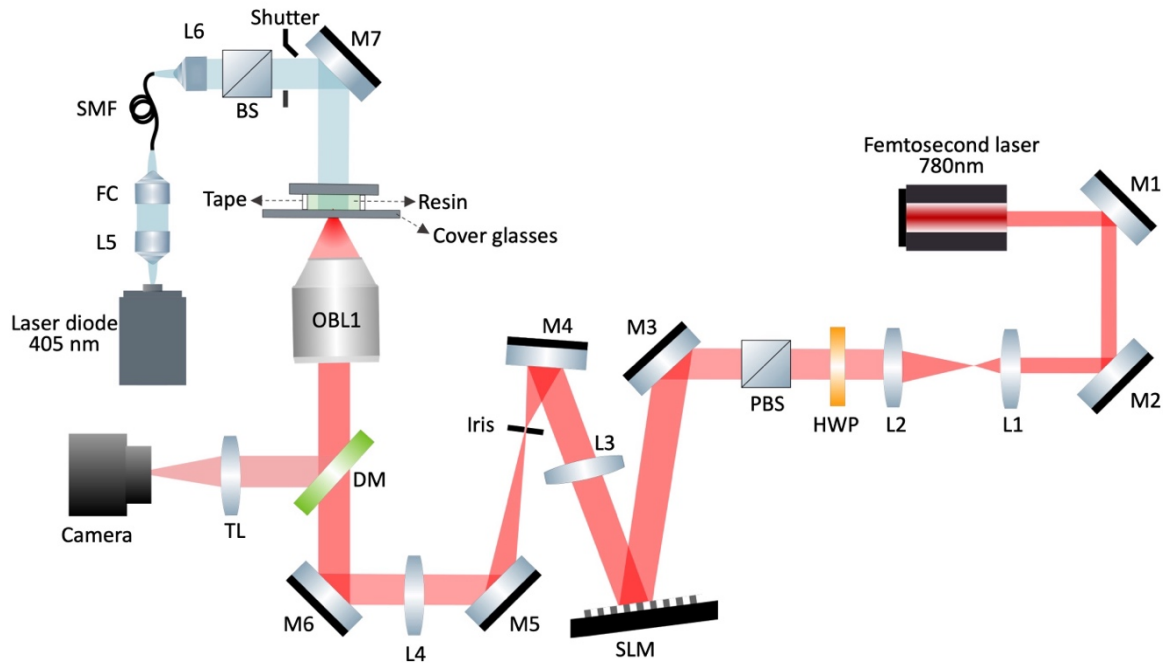

**Figure S. 1.** Illustration of the experimental setup for blue light illumination-assisted two-photon polymerization. The light path with red color represents the 2PP part while the blue color light path is for the 1PA process.

## 2. Characterization of SLM response time

When an electric field is set across each pixel electrode in the SLM, the orientation of the LCoS and thus the light retardation changes in proportion to the field. Hence, the shape of the incoming plane wavefront is modified, i.e., phase modulation (retardation). However, the reorientation of the LCoS requires a certain time to react and stabilize. The rise and fall times are referred to as SLM response time, and generally do not exceed the input frame rate of SLM [1]. In order to determine the response time of the SLM we used in the experiment, a beam spot is moved from one position to another position (in Fig. S.2.a) by changing the phase pattern/mask on the SLM. 400 frames per trigger are collected where the change in the intensity for the maximum pixel coordinates of the first frames (reference) is recorded over the measurement as seen in Fig. S.2.b. While switching from position 1 to position 2 by applying different phase masks, the effects of the fall time and rise are observed clearly in Fig. S.2.b. The response time of the LCoS is measured to be around 40 ms, which means that the time for each successive focal spot is limited by this SLM response time. For printing, a margin of 50 ms is employed to ensure precision in timing. Additionally, the pulse width modulation method for digital addressing of pixels causes fluctuating voltage response in SLM [2]. Therefore, flickering in intensity is observed.

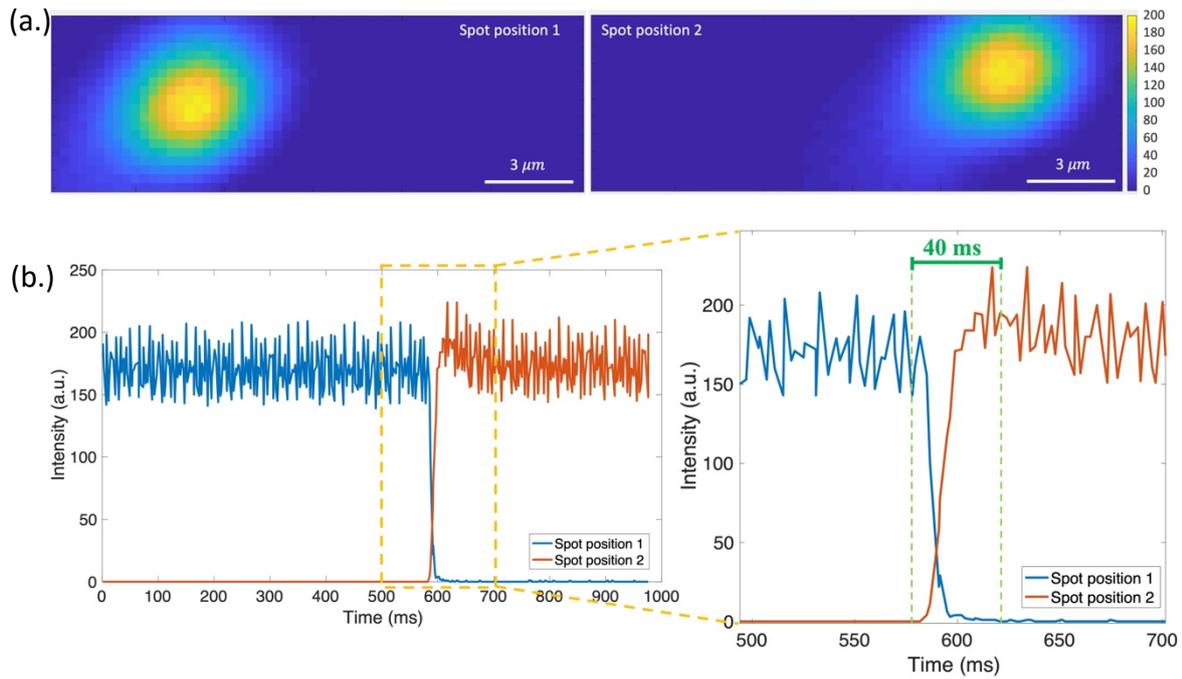

**Figure S. 2.** SLM Response time characterization. (a.) Beam spot at position 1 (left) and position 2 (right). (b.) Change in the relative intensity while switching from position 1 to position 2 by varying the period of the phase patterns. Response time is measured in the interval of falling time for position 1 and rising time for position 2 as around 40 ms.

### 3. Dose test for blue light illumination

The response of the resin to the blue light illumination is characterized by irradiating the resin with several blue light doses where the average blue light laser power and exposure time are varied. After rinsing the unpolymerized resin using PGMEA and IPA respectively, the polymerized region is imaged by a PC microscope. The diameters of the printed area are measured and plotted as a function of the applied light dose as shown in Fig. S.3. The threshold dose is determined as 8.02 mJ/cm$^2$ which is the highest dose level before solidifying the resin.

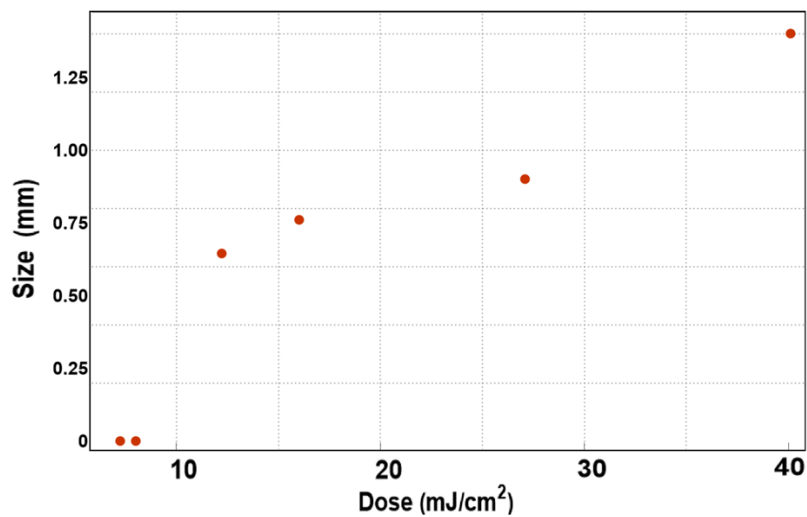

**Figure S. 3.** Graph of the dose test for blue light illumination.

## 4. Dose test for blue light illumination-assisted two-photon polymerization

An array of focused spots is printed by varying the average fs light power: 20,23,25, and 30 mW and the exposure time of the beam spots is varied (from 50 ms to 3 s) along each row of the array. The experiment is repeated 4 times. The spot arrays are then imaged by a phase contrast (PC) microscope without dissolving the surrounding uncured resin. The polymerized spots have a different refraction index than the surrounding liquid resin and thus can be visualized by the PC microscope. The diameters of the printed voxels are measured and plotted as a function of the applied light dose as shown in Fig. S.4, fitted with a logarithmic transform function. As expected, the printed voxel size decreases when the average power and the exposure time are reduced. When the power is set below 20 mW, we could not obtain a contrasted image of the printed voxels, likely because the spot sizes and the contrast are below the resolution of the PC microscope (40x/0.60 objective lens).

We then proceed with a characterization of the resin response to both blue light illumination and fs illumination. The resin is first exposed to a dose of 7.7 mJ/cm$^2$ blue light (corresponding to an illumination power of 15 mW for 100 ms) over an area of 0.193 cm$^2$, followed by a fs laser beam exposure. We first observed that with 18 mW fs power, the printed voxels can be discerned with the PC microscope, whereas it was not the case when the same fs power was used without blue light. In Fig S.4. b, the voxel size is considerably larger compared to the situation when only the fs beam is employed.

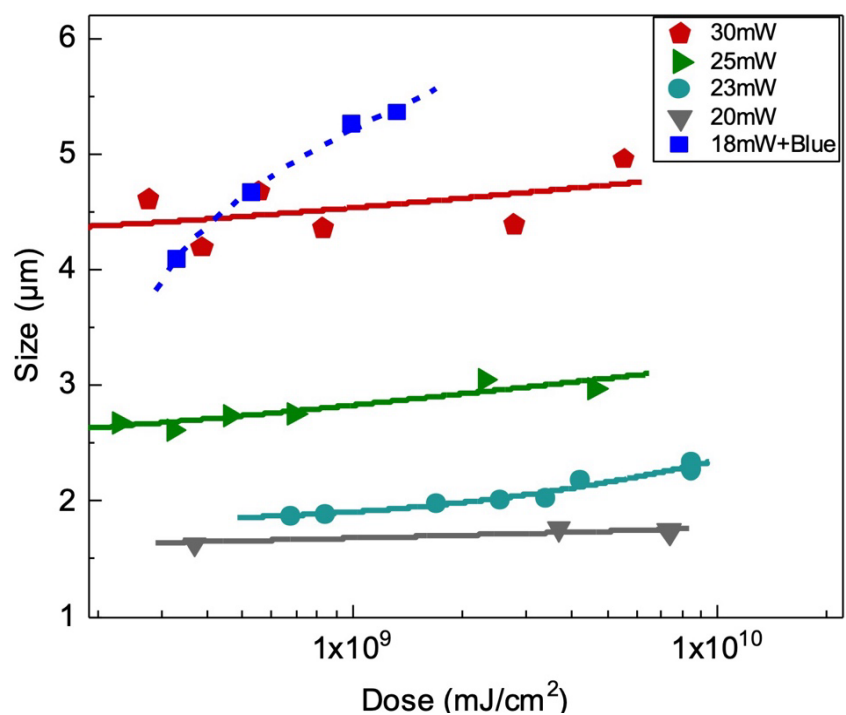

**Figure S. 4.** Dose test with fs light powers of 30, 25, 23, and 20 mW; 18 mW assisted by blue light illumination. Voxel sizes are measured without rinsing the unsolidified resin surrounding the prints.

## 5. Experimental setup for blue light-sheet assisted two-photon polymerization

The light-sheet-assisted two-photon polymerization setup is illustrated in Fig. S.5. The only modification for the 2PP part represented in Supplementary Note 1 is OBL1 which is replaced with a low NA objective lens (Zeiss A-Plan 20x, NA = 0.45), while other components remain unchanged. A glass rod (Hilgenberg GmbH, made of borosilicate glass 3.3, ends cut L= 35 Diameter = 0.7 ± 0.03 mm) is employed as a printing substrate, for which a coated coverslip (700x700 µm$^2$) is glued by an optical adhesive onto the tip of the rod to ensure and reinforce the adherence of the printed parts to this substrate. To monitor the printing process, a red LED passing through the rod and the camera with an IR blocking filter (Thorlabs TF1) are integrated. Resin is placed into a cuvette (ibidi GmbH, µ-Slide 8 Well high Glass Bottom #1 1/2).

A blue light-sheet system is integrated into this modified 2PP setup. The laser diode at 405 nm wavelength is coupled into the SMF as described in Supplementary Note 1, and the output of the SMF is mounted to a fiber collimator (FC2, AC080-016-A-ML f = 16 mm). A beam splitter (BS, Thorlabs BS031 50:50) is used on the light path to measure the incidence power, and the optical beam shutter is placed on the light path to control the exposure time of the blue light sheet as well. A cylindrical achromatic doublet (CYL, ACY254-050-A f = 50 mm) with a low NA objective lens (OBL2, Olympus Plan N 4x, NA = 0.10) is integrated to generate the blue light sheet with a waist of 3.4 µm.

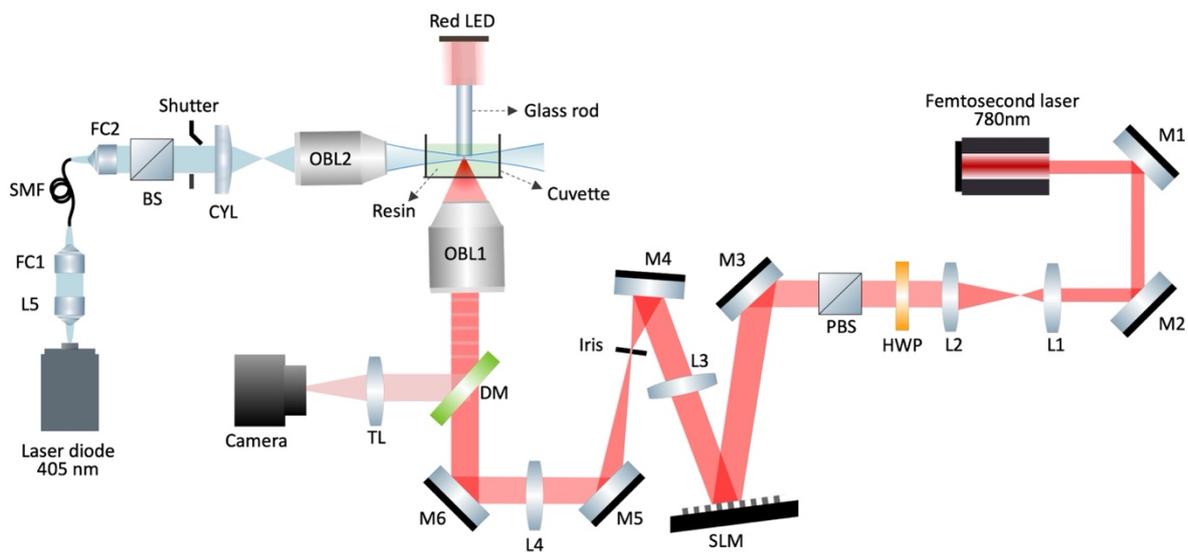

**Figure S. 5.** Illustration of the experimental setup for blue light-sheet assisted two-photon polymerization. The light path with red color represents the 2PP part while the blue color light path is for the 1PA process by the light-sheet implementation.

## 6. Blue light-sheet characterization

Blue light-sheet is characterized using a camera placed in front of the objective lens OBL2 in Fig. S.6. The camera is mounted on a motorized stage, and the spatial shape of the beam is captured with a 50 µm step size along a 12 mm path. Stacks of the collected two-dimensional (2D) images are demonstrated in Fig. S.6.a. Corresponding beam waist $w_0(z)$ as a function of z-axis direction is given in Fig. S.6.b. Gaussian fit is applied to the collected data for retrieving the information on the beam waist as depicted in Fig. S.6.c with an inset featuring a sample of collected 2D images close to the focal region. The waist of the blue light-sheet at the focal region is obtained as 3.4 µm (5.02 µm) yielding a Rayleigh range of 89.7 µm (289 µm) and a lateral (xz plane) beam expansion of 3.42 mm (5.62 mm) in the air (inside the resin with 1.48 refractive index).

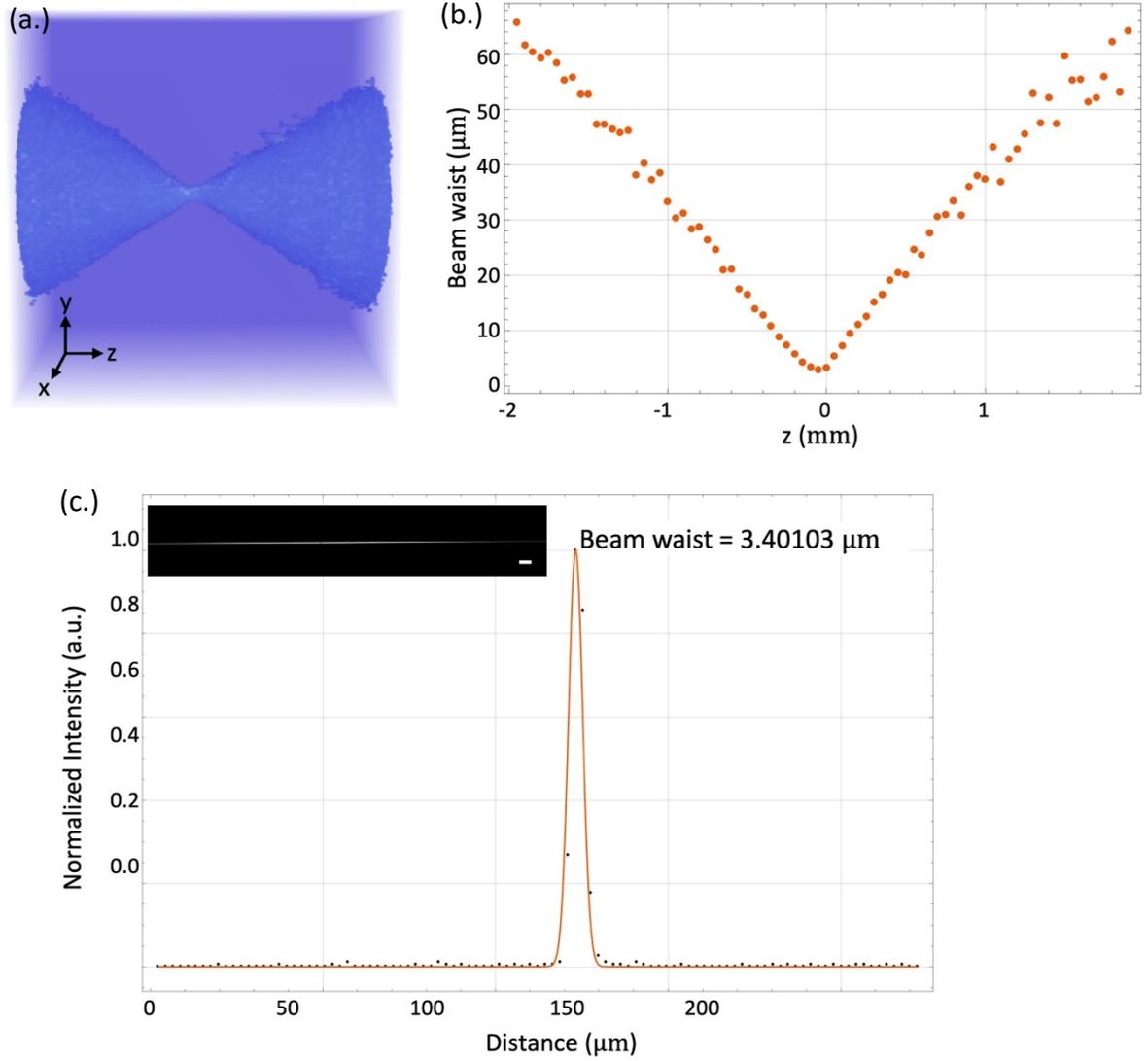

**Figure S. 6.** Blue light-sheet characterization. (a.) Three-dimensional (3D) reconstruction of the blue light-sheet, and (b.) corresponding beam waist w0(z) as a function of z-axis direction. (c.) Beam profile at the focus of the beam (beam waist, w0) with an inset featuring a sample of collected 2D images close to the focal region. The scale bar is 50 μm.

## 7. Digital control of beam spot position

The phase-only SLM provides greater flexibility to the system by offering digital control on the beam, rather than relying on a galvo mirror commonly used for lateral beam scanning in the x and y directions. To steer the beam, phase gratings are displayed on the SLM for two-dimensional control of the beam spot on the print plane. In detail, the phase distribution used to scan the beam is given by [3,4]:

$$\varphi_{grat}(x,y) = \left(\frac{2\pi}{\Lambda_x}x + \frac{2\pi}{\Lambda_y}y\right) \qquad (1)$$

Accordingly, the linear shift in the printing plane is generated by changing the periods $\Lambda_x$ and $\Lambda_y$, which are the grating periods in the x- and y-directions, respectively. To perform beam steering, pre-computed phase gratings with different periods (examples are given in Fig. 1.a) are sequentially displayed on the SLM. After the display time has elapsed, the glass rod is synchronized to move up towards the z-direction using the motorized linear stage to start a new layer adjusting to the previously printed layer. The axial step size (slicing) that the glass rod moves is highly dependent on the beam power since various powers result in different voxel heights. Yet, bypassing the need for mechanical rod movement to gather the layers, Fresnel lenses can be introduced to the system via SLM for a digital precision in the axial beam control [5,6]. The beam focus is translated in z-axis, either in the positive or negative direction using the phase distribution given by:

$$\varphi_{Len}(x,y,z) = \frac{2\pi z}{\lambda f^2}(x^2 + y^2) \qquad (2)$$

Where $f$ is the focus of the Fourier lens, $\lambda$ is the wavelength of the light, and $z$ is the axial shift relative to the focal distance. The superposition of phase elements (1) and (2) is performed to create a stack of holograms that controls the position of the beam spot in three dimensions.

$$\varphi^{3D}(x,y,z) = \left[\varphi_{grat}(x,y) + \varphi_{Len}(x,y,z)\right] \qquad (3)$$

**References**

[1] Spatial Light Modulators - HOLOEYE Photonics AG 2023. https://holoeye.com/products/spatial-light-modulators/ (accessed January 3, 2024).
[2] Zheng M, Chen S, Liu B, Weng Z, Li Z. Fast measurement of the phase flicker of a digitally addressable LCoS-SLM. Optik 2021;242:167270. https://doi.org/10.1016/j.ijleo.2021.167270.
[3] Haist T, Osten W. Holography using pixelated spatial light modulators—part 1: theory and basic considerations. J MicroNanolithography MEMS MOEMS 2015;14:041310. https://doi.org/10.1117/1.JMM.14.4.041310.
[4] Rosales-Guzmán C, Forbes A. How to Shape Light with Spatial Light Modulators. SPIE PRESS; 2017. https://doi.org/10.1117/3.2281295.
[5] Curtis JE, Koss BA, Grier DG. Dynamic holographic optical tweezers. Opt Commun 2002;207:169–75. https://doi.org/10.1016/S0030-4018(02)01524-9.
[6] Melville H, Milne G, Spalding G, Sibbett W, Dholakia K, McGloin D. Optical trapping of three-dimensional structures using dynamic holograms. Opt Express 2003;11:3562. https://doi.org/10.1364/OE.11.003562.